\documentclass[prd,twocolumn,superscriptaddress,floatfix]{revtex4-1}
\usepackage[utf8]{inputenc}
\usepackage{graphicx}
\usepackage{graphics,bm}
\usepackage{amssymb}
\usepackage{color}

\newcommand{\ket}[1]{| #1 \rangle}

\def\6{\langle}
\def\9{\rangle}

\usepackage{amsmath}
\usepackage{amsfonts}
\usepackage{amssymb}

\newcommand\wtPsi{\widetilde{\mathtt{\Psi}}}
\newcommand\wtG{\widetilde{\mathtt{G}}}
\newcommand\wtPi{\widetilde{\mathtt{\Pi}}}
\newcommand\wtu{\widetilde{\mathtt{u}}}
\newcommand\wta{\widetilde{\mathtt{a}}}

\newcommand{\be}{\begin{equation}}
\newcommand{\ee}{\end{equation}}
\newcommand{\ba}{\begin{eqnarray}}
\newcommand{\ea}{\end{eqnarray}}

\usepackage{amsmath}

\begin{document}
%opening
\title{On the nature of quantum gravity}

\author{Vasileios I. Kiosses}
\email{kiosses.vas@gmail.com}
\affiliation{Instituto de F\'{i}sica, Universidade Federal de Goi\'{a}s, Caixa Postal 131, 74001-970, Goi\^{a}nia, Brazil}

\begin{abstract}
It was recently advanced the argument that Unruh effect emerges from the study of quantum field theory in quantum space-time. Quantum space-time is identified with the Hilbert space of a new kind of quantum fields, the accelerated fields, which are defined in momentum space. In this work we argue that the interactions between such fields offer a clear distinction between flat and curved space-times. Free accelerated fields are associated with flat spacetime, while interacting accelerated fields with curved spacetimes. Our intuition that quantum gravity arises via field interactions is verified by invoking quantum statistics. Studying the Unruh-like effect of accelerated fields, we show that any massive object behaves as a black body at temperature which is inversely proportional to its mass, radiating space-time quanta. With a heuristic argument it is shown that Hawking radiation naturally arises in a theory in which space-time is quantized. Finally, in terms  of thermodynamics, gravity can be identified with an entropic force guaranteed by the second law of thermodynamics.
\end{abstract}

\maketitle

%%%%%%%%%%%%%%%%%%%%%%%%%%%%%%%%%%%%%%%%%%%%%%%%%%%
\noindent

\section{Introduction}

Quantum field theory and general relativity are considered as the most successful theories we have, however they are based on contradictory hypotheses. General relativity teach us that space-time is curved and everything is smooth and deterministic. On the other hand, quantum field theory advocates that the world is formed by discrete quanta over a flat space-time, governed by global symmetries. 

As the history of physics has shown, the comparison between apparently contradictory successful theories has led to major steps in science. Accelerated quantum fields \cite{C-K1,C-K2} are precisely an effort to solve the irreconcilable contradiction between quantum mechanics, as formulated in quantum field theory, and general relativity. In this approach, both the lessons of geometry and quantum are taken into account and we proceeded with a fresh look at the problem. 

In accelerated quantum field theory, accelerated reference frames are represented by quantum space-time. It is usually assumed that space-time should be a continuum in order to define a quantum field theory in an accelerated frame of reference. This is due to the fact that the transformation of a quantum field to an accelerating frame is simply implemented by the coordinate transformation from inertial frame to an accelerating one. The main purpose of \cite{C-K1} was to show that an appropriate modification of the ordinary concept of space-time redefines an accelerating quantum system of fields. The main result in that work is that there exists a Lorentz invariant definition of space-time that contains a natural unit of length which depends on the value of acceleration. In a subsequent paper, which was aiming at demonstrating the utility of accelerated fields, it was shown how Unruh effect emerges from quantum space-time \cite{C-K2}.

The incorporation of gravity into relativistic framework may be based on the equivalence of all accelerated systems. Einstein's important advance was to realize that if all accelerated systems are equivalent, then Euclidean geometry cannot hold in all of them. Thus, Riemannian geometry has been manifested itself as the natural mathematical framework to study gravity. Accelerated quantum fields modify the concept of accelerating reference frame and therefore our understanding of gravity. The description of gravity as space-time curvature cannot survive in this theory. Our aim in this work is to show how we can understand gravity in the theory of accelerated quantum fields. As we will see, gravity appears as a feature of accelerated fields, which is necessary for the consistence of the theory, thus unraveling its emergent nature.

On the way to quantum gravity, while it is generally accepted that space-time is quantized, there is disagreement as to how quantization manifests itself \cite{Smolin}. In accelerated quantum field theory, the definition of quantum fields in momentum space indicates the quantization of space-time in a mathematically consistent way. Space and time are quantized in the way quantities like energy and momentum are quantized in ordinary quantum field theories. 

The requirement of locality remains a strong motivation for studying field theories in the quantum world. However, the presence of a fundamental length scale in any theory of quantum gravity guaranties the entrance of non-locality. Therefore we cannot treat the quantum gravitational field simply as a quantum field in space. Physicists have sought a way to incorporate gravity into a quantum field theory by making conjectures about possible alterations that could be made to the theory. Maldacena suggested that holography might be the key to reconcile gravity with quantum mechanics, all you need is an extra dimension of space [AdS/CFT] \cite{Maldacena}. Our proposition renders extra dimensions needless. We share Born's view \cite{Born}, that physics should be equivalently formulated from the position and momentum point of views, and develop a theory where space-time and momentum space appear to hold equal parts. We promote momentum space by constructing a relativistic field theory in momentum space analogously to matter field theories in space-time. As a consequence, in our case, the smooth metric geometry of space, which is the ground needed to define a quantum field, is provided, not by the space-time but, by the momentum space.

Constructing a relativistic field theory in momentum space analogously to standard field theories in space-time, it would be interesting to investigate whether and in which form the properties which characterize standard quantum field theories appear in the new construction of quantum fields. We find that Unruh effect \cite{Unruh}, one of the most intriguing feature of quantum field theory, where different frames of reference correspond to different vacua, is also met in accelerated field theory, with the exception now that the key-role is on mass and not on acceleration.  

Invoking equivalence principle, we apply accelerated field theory to uniform gravitational field and show that the corresponding Unruh ambiguity in the vacuum definition is the quantum analogue of tidal effects of gravity. In addition, we are led to a new understanding of Hawking radiation, which simply appears to be the Unruh effect of accelerated fields.
In this way, gravity is represented by a type of black body radiation, which is emitted by any massive object. Instead of energy quanta, a ``gravitational'' black body emits quanta of space-time. Alternatively, this can also be expressed in the following way: The central idea of general relativity, that matter causes space-time to curve, in a field theory, is translated to the fact that all massive objects emit quanta of accelerated fields.

The point that gravity, and space time, can indeed be explained in terms of quantum field terms, may have important implications not only regarding the foundations of physics, but also for many areas in which gravity plays a central role.

In section II we briefly summarize the main elements of accelerated field theory. Section III illustrates how curved space-times arise in the framework of accelerated quantum field theory. In section IV we study the Unruh-like effect of accelerated fields and explain how Hawking radiation can be derived, while section V is devoted to discuss the origin of mass in a context in which space is quantized. We close with an overview of the results and some discussions in section VI.

Throughout the text we consider quantum field theories in two dimensions with metric signature $(+,-)$. Furthermore, the units are chosen such that $c = 1$ but $\hbar$ and $G$ are kept explicit.

\section{spacetime as a quantum object}
There is plenty of evidence today for physicists to come to a consensus
supporting an all-fields view. However, the conventional intuition provided by quantum field theory fails for quantum gravity. The view where quantum fields are defined over space-time, as a continuum, needs to be abandoned for quantum gravity. We are in need of a novel way to define space and time. This is exactly what accelerated quantum fields theory \cite{C-K1,C-K2} provides. In this section, we review the main features of accelerated quantum fields, a recent development of a quantum description of space-time, done in terms of a field theory. 

The point of departure is Unruh radiation, one of the most interesting effects in modern physics. According to Unruh \cite{Unruh}, an accelerated observer moving through Minkowski vacuum will see a background of quanta, but an inertial observer will swear that the state is empty. Accelerated quantum field theory opens the possibility of an alternative explanation. More specifically, let us consider the simple case of a real massive bosonic quantum field which obeys the Klein-Gordon equation $\left(\partial_t^2 - \partial_x^2 +m^2\right) \Phi =0$. Instead of taking the Unruh-Fulling approach, in which the field equation defined in an accelerated frame is expressed in Rindler coordinates, we introduce a field in momentum space, defined by the wave equation 
\be
\left({\widetilde{\partial}_E}^2-{\widetilde{\partial}_p}^2 - 1/\alpha^2 \right)\wtPsi(E,p)= 0.\label{eq.1}
\ee
Then, the thermal spectrum arises mixing the two fields   
\be
\left\{
\begin{aligned}
	\left(\partial_t^2 - \partial_x^2 + m^2\right) \Phi(t,x) &=0 \\
	\left({\widetilde{\partial}_E}^2-{\widetilde{\partial}_p}^2 - \frac{1}{\alpha^2} \right)\wtPsi(E,p) &= 0
\end{aligned}
\right.
\ee
For more details I refer to \cite{C-K2}.	

Eq.(\ref{eq.1}) promotes the relativistic kinematic relation $x^2-t^2=1/\alpha^2$ to field status. But, what is the physical meaning of a field which is defined in momentum space? To answer this it would help to recall that, according to general relativity, physical processes associated with the gravitational field occur even in space that is free from matter fields. Obviously, $\wtPsi(E,p)$ survives in a physical system with no matter field. So, even if $\wtPsi(E,p)$ not stands directly for the gravitational field, it should be associated in a way with space-time (the quantity that remains even with no matter fields).

The modern view we have for fields is that they are states or conditions of space. Accelerated fields theory generalizes this view by introducing the ``states'' or ``conditions'' of momentum space. In this context, $\wtPsi(E,p)$ remains continues, filling all the momentum space, while quantizing it, ``space'' and ``time'' become the primary dynamical concepts, with acceleration to be a parameter that appears in the description of ``space-time''.

A straightforward approach to the quantization of the free field $\wtPsi(E,p)$ begins by expanding the field in an orthonormal set of mode solutions $\wtu_t(E,p) = e^{i(E t - p x_t)}/\sqrt{4 \pi x_t}$, with $x_t \equiv x(t) = \sqrt{t^2 + 1/\alpha^2}$, and claiming that the definition of the positive- and negative-frequency solutions lies in the existence of a space-like Killing vector field, $\widetilde{\partial}_p$, in momentum space \cite{C-K1}. Working in the space spanned by the positive frequency modes $\wtu_t$, one can define the field operator $\wtPsi$ in the language of the second quantization in the usual way
\be
\wtPsi(E,p) = \int dt \left( \wta_t \wtu_t(E,p) + \wta_t^\dagger \wtu_t^*(E,p)\right).
\label{eq.FOM}
\ee
$\wta_t$ and $\wta_t^\dagger$ are the annihilation and creation operators, respectively, since they fulfill the typical algebra for creation and annihilation operators, i.e. 
\be
\left[\widetilde{\mathtt{a}}_{t},\widetilde{\mathtt{a}}_{t'}\right] = \left[\widetilde{\mathtt{a}}_{t}^\dagger,\widetilde{\mathtt{a}}_{t'}^\dagger\right] = 0,\quad \left[\widetilde{\mathtt{a}}_{t},\widetilde{\mathtt{a}}_{t'}^\dagger\right] = \delta(t-t') \label{eq.61}
\ee
By defining the conjugate momentum as $\wtPi(E,p) = \widetilde{\partial}_p \wtPsi(E,p)$, the commutation relations (\ref{eq.61}) in space-time are equivalent to the canonical \emph{equal-momentum} commutation relations
\ba
\left[\wtG_p(E),\wtG_p(E')\right]&=& \left[\wtPi_p(E),\wtPi_p(E')\right] =0, \nonumber %\label{crmI}
\\
\left[\wtG_p(E),\wtPi_p(E')\right]&=& i \, \delta(E-E'),\nonumber %\label{crmII},
\ea
in momentum space.

Which is the spectrum of the theory? To begin with, we define the Hamiltonian of the theory. The Hamiltonian is defined in the following way. First we construct the Lagrange density associated with equation (\ref{eq.1}) by inverting the Euler-Lagrange equation. Then, Noether's theorem provides us the momentum independent quantity that plays the role of the Hamiltonian. Finally, by using Eq. (\ref{eq.FOM}) we can write the Hamiltonian in the form
\be
\widetilde{X} = \int \frac{dt}{\sqrt{4\pi x_t}} \,  x_t\, \left(\widetilde{\mathtt{a}}_{t}^\dagger\,\, \widetilde{\mathtt{a}}_{t} +\frac{1}{2}\left[\widetilde{\mathtt{a}}_{t},\widetilde{\mathtt{a}}_{t}^\dagger\right]\right).\label{Ham-pos}
\ee
The second term inside parenthesis is something we cannot avoid. Our treatment resembles that of harmonic oscillator and this term is the field analogue of the harmonic oscillator zero-point energy. We will ignore this term in all of our calculations below, unless stated otherwise. But, we intend to discuss the nature and the consequences of this term in a subsequent work. From Noether's theorem it also follows the operator (see \cite{C-K1} for details)
\be
\widetilde{T} = - \int dE\,\, \widetilde{\mathtt{\Pi}} \widetilde{\partial}_E \widetilde{\mathtt{G}} = \int \frac{dt}{\sqrt{4 \pi x_t}} \,  t\,\,\widetilde{\mathtt{a}}_{t}^\dagger\,\, \widetilde{\mathtt{a}}_{t} . \label{T-time}
\ee 
Using Eq. (\ref{Ham-pos}) for the Hamiltonian, it is straightforward to evaluate the commutators
\be
\left[\widetilde{X},\widetilde{\mathtt{a}}_{t}^\dagger\right] = x_t \widetilde{\mathtt{a}}_{t}^\dagger,\qquad \left[\widetilde{X},\widetilde{\mathtt{a}}_{t}\right] = -x_t \widetilde{\mathtt{a}}_{t}.
\ee
We can then write down the spectrum of the theory. There will be a single vacuum state $\ket{0_{\alpha}}$, characterized by the fact that it is annihilated by all $\widetilde{\mathtt{a}}_t$,
\be
\widetilde{\mathtt{a}}_t \ket{0_{\alpha}} =0,\qquad  \forall t.
\ee 
All other eigenstates can be built by letting $\widetilde{\mathtt{a}}_t^\dagger$ acting on the vacuum,
\be
\ket{t} = \widetilde{\mathtt{a}}_t^\dagger \ket{0_{\alpha}}.
\ee
Acting on our $\ket{t}$ with the operators $\widetilde{X}$ and $\widetilde{T}$ we obtain the quantum numbers of the state.
\ba
\widetilde{X} \ket{t} &=& x_t \ket{t}\qquad \text{with}\qquad x_t^2 = t^2 + 1/\alpha^2 \\
\widetilde{T} \ket{t} &=& t \ket{t}.
\ea
As we see the quantum numbers of the state are related to space and time. Hamiltonian $\widetilde{X}$ has dimensions of length, while $\widetilde{T}$ has dimension of time.
  
Before proceed to the interpretation of the eigenstates $\ket{t}$, let us recall how an accelerated frame appears in special relativity and what is the physical meaning of the coordinates $(x_t,t)$ appearing in the equation $x_t^2 = t^2 + 1/\alpha^2$. Consider an accelerated observer with constant acceleration $\alpha$. Relative to an inertial frame, his motion is given by the world line $x^\mu=x^\mu(\tau)$ ($\tau$ is the proper time). The world velocity $u^\mu=dx^\mu/d\tau$ is a time-like unit vector and because it is fixed in magnitude it is orthogonal to the world acceleration $a^\mu=du^\mu/d\tau$. Joining to these conditions the equation $a^\mu a_\mu=\alpha^2$, the solution of the equations for the motion of the observer in the inertial frame reads \cite{MTW} $x^\mu x_\mu\equiv x^2 - t^2 = 1/\alpha^2$. The world line of a uniformly accelerated observer in a space-time diagram is a hyperbola. For each coordinate time $t$ the position of the observer is
\be
x= \sqrt{t^2 + 1/\alpha^2},
\ee 
which is equivalent to the eigenstate $x_t$.

The theoretical concepts that physicists have formed about space and time, the geometric theory of gravity among them, are based on the assumption that the variables $x_t$ and $t$ (if we restrict ourselves to $1+1$ space-time) take on a continuum of values and they may take on these values simultaneously. If this constitute the classical aspect of space and time, the quantum is introduced by postulating that space and time are composed of discrete quanta. We theorize that the position $x_t$ may be equal to an angular frequency $\widetilde{\omega}$, while time $t$ may be equal to a wavenumber $\widetilde{k}$, both multiplied by a constant $\hbar$,
\be
x _t = \hbar \, \widetilde{\omega}_t\qquad \text{and}\qquad t = \hbar \, \widetilde{k},
\ee
following the recipe of one of the starting milestones of quantum theory, the wave-particle duality. Notice that in this setting  $\hbar$ has dimensions of physical action thus it is nothing else than the reduced Planck constant. Equation (\ref{eq.1}) has been developed principally from the above hypothesis, a wave equation that would describe quantum space-time. In practice, natural units comprising $\hbar=1$ are used, except otherwise stated, allowing time, wavenumber, length and angular frequency to be used interchangeably. Thus, we recognize the relation $x_t^2 = t^2 + 1/\alpha^2$ as the relativistic dispersion relation for a frame of reference in a state of acceleration $\alpha$.

To this end, let us try to interpret the eigenstates of the Hamiltonian, the states of the quantum field $\wtPsi$. The field equation for $\wtPsi$ together with eq.(\ref{eq.FOM}) imply that the amplitudes $\wta_t$ and $\wta^\dagger_t$ of the $t_{\text{th}}$ field mode satisfy the equations of motion for a set of quantum harmonic oscillators. These equations should be seen as the equations of motion for a mechanical system having not one, but an infinite number of degrees of freedom. Operators $\wta_t$ and $\wta^\dagger_t$ are the familiar raising and lowering operators from the harmonic oscillator problem. As in the harmonic oscillator problem, the $t_{\text{th}}$ mode has an infinite “discrete” position spectrum $x_t(n_t+1/2)$. $n_t=0,1,2,\cdots,\infty$ is the number of quanta in the $t_{\text{th}}$ mode and is the eigenvalue of the number operator $\widetilde{N}_t :=\widetilde{\mathtt{a}}_{t}^\dagger\,\, \widetilde{\mathtt{a}}_{t}$. Notice that we have to keep track of separate numbers of quanta for each time $t$. A distinctly quantum aspect is that, even in the vacuum state where $n_t= 0$, each mode has length $x_t/2$. Another quantum aspect is that the position $x$ of an accelerating observer of a single mode has an infinite spectrum of discrete values separated by $\Delta x = x_t$.
The first excited state $\ket{t}$, as we saw after dropping the infinite constant, has length $x_t$ and time interval $t$. We interpret $\ket{t}$ as the spacetime eigenstate of a single accelerating observer of acceleration $\alpha$ with coordinates $(x_t,t)$.
Each quantum of the field is called an excitation, because its length $x_t$ represents more "lengthy" accelerating observers.
All other spacetime eigenstates can be built by acting on $\ket{0_{\alpha}}$ with creation operators. In general, the state 
\be
\ket{t_1 +t_2+\cdots} = \widetilde{\mathtt{a}}_{t_1}^\dagger \,\widetilde{\mathtt{a}}_{t_2}^\dagger\cdots \ket{0_{\alpha}} \label{eq.14}
\ee
is an eigenstate of $\widetilde{X}$ with length $x_{t_1}+x_{t_2}+\cdots$ and of $\widetilde{T}$ with time interval $t_1+t_2+\cdots$. 

Equation (\ref{eq.14}) shows that accelerated field theory can accommodate multi-quanta states. We interpret the state in which $n$ 
$\widetilde{\mathtt{a}}_t^\dagger$ acts on the vacuum as an $n$-accelerating observers state. 
The full Hilbert space of our theory is spanned by acting on the vacuum with all possible combinations of $\widetilde{\mathtt{a}}_{t}^\dagger$'s,
\be
\ket{0_{\alpha}},\quad \widetilde{\mathtt{a}}_{t_1}^\dagger \ket{0_{\alpha}},\quad \widetilde{\mathtt{a}}_{t_1}^\dagger \,\widetilde{\mathtt{a}}_{t_2}^\dagger \ket{0_{\alpha}}, 
\quad \widetilde{\mathtt{a}}_{t_1}^\dagger \,\widetilde{\mathtt{a}}_{t_2}^\dagger\,\widetilde{\mathtt{a}}_{t_3}^\dagger \ket{0_{\alpha}},\, \cdots
\ee
This space is the corresponding Fock space of the theory. A useful operator which counts the number of accelerated observers in a given state in the Fock space is the number operator 
\be
\widetilde{N} :=  \int \frac{dt}{\sqrt{4 \pi x_t}} \,  \,\,\widetilde{\mathtt{a}}_{t}^\dagger\,\, \widetilde{\mathtt{a}}_{t}.
\ee
and satisfies $\widetilde{N} \ket{t_1+\cdots + t_n}=n \ket{t_1+\cdots + t_n}$. 

It is important to note that the number operator commutes with the Hamiltonian, $[\widetilde{N},\widetilde{X}] = 0$, ensuring that particle number is conserved. This is a property of free theories where, accelerated observers do not "interact" with each other. However, this will no longer be true in the rest of the paper where we consider interactions: interactions create and destroy accelerating observers, taking us between the different sectors in the Fock space. In the sections that follow we argue that exactly this transition between different sectors in the Fock space identifies gravity.  

\section{free-fall observers in accelerated field theory}

In Newtonian physics, free fall is any motion of a body where gravity is the only force acting upon it. In the context of general relativity, where gravitation is reduced to a space-time curvature, a body in free fall is subject to no force and is an inertial body moving along a geodesic.

Due to equivalence principle, an infinitesimally extended, homogeneous gravitational field can be completely replaced by a state of acceleration of the reference system. In the framework of accelerated quantum fields theory, equivalence principle permits us to identify the world line of a free fall observer to the Hilbert space of an accelerated field where the parameter of inertial acceleration is replaced by the gravitational acceleration. More specific, let us consider a real field $\wtG(E,p)$ defined at all points $(E,p)$ of momentum space, satisfying the field equation
\be
\left({\widetilde{\partial}_E}^2-{\widetilde{\partial}_p}^2 - \frac{1}{g^2} \right)\wtG(E,p) = 0,
\label{eq.wem}
\ee
where $g$ is to be interpreted as the gravitational acceleration of the field quanta.
Following the formulation of accelerated quantum field theory, the field operator has the expansion
\be
\wtG(E,p) = \int dt \left( \wta_t \wtu_t(E,p) + \wta_t^\dagger \wtu_t^*(E,p)\right).
\label{eq.FOM-2}
\ee
In addition, our discussion at the previous section shows that the vacuum state is defined by $\widetilde{\mathtt{a}}_{t} \ket{0_g}=0$. The states $\widetilde{\mathtt{a}}_{t}^\dagger\ket{0_g}$ having time interval $t$ and length $x_t$ are interpreted as single free falling observer states at position with coordinates $(t,x_t)$. 
States of the form $\widetilde{\mathtt{a}}_{t_1}^\dagger \,\cdots \widetilde{\mathtt{a}}_{t_n}^\dagger \ket{0_g}$ are interpreted as $n$-free falling observers states.

Up to this point, we have seen no interactions. The free field theory that we have discussed so far is very special: we can determine its spectrum, but nothing interesting then happens. It has field excitations, but these excitations do not interact with each other.
Combining more than two accelerated fields, i.e. fields whose parameter $g$ takes different values, typically yields a quantum operator which changes the particle number. The interaction terms in the Hamiltonian can be used to create free fall observers with new coordinates or annihilate observers leaving behind length only in the form of other free fall observers. Thus, the number of degrees of freedom becomes a dynamical variable.

To understand what is at stake, we need to look back at the precedents, where spacetime is described geometrically, and recall a fundamental implication of gravity: the tidal distortions of free fall observers.

\begin{figure}[htbp]
	\includegraphics[width=0.45\textwidth]{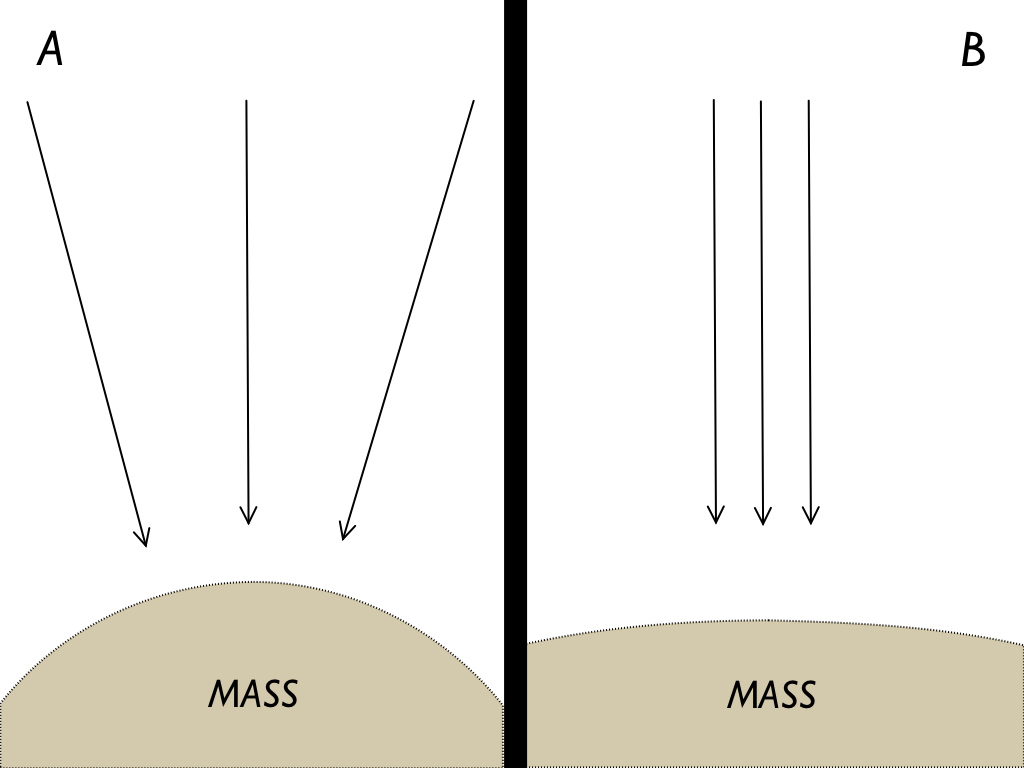}
	\caption{Free fall accelerations for test particles released along an axis in the plane tangent to the mass's surface. In A: The trajectories, which are directed to the center of the mass, are in different directions. Thus, there is a relative acceleration among the particles. In B: Point particles arranged in a small distance, no relative acceleration is detected. 
	}
	\label{figure}
\end{figure}

Consider a free falling observer in a massive object's gravitational field (see figure \ref{figure}). The observer places three point particles along an axis in the plane tangent to the massive object’s surface. The released point particles are independent in the sense that, once placed, they are in free fall. There are no external forces acting on them except gravity, so they are not moving relative to each other. All three point particles point to the center of mass, but depending on how far between each other are placed, they have their free fall accelerations directed differently or not. These two situations are depicted in figure \ref{figure}.
In case the point particles are not falling in precisely the same direction (figure \ref{figure}-A), although the magnitudes of the free fall accelerations are the same, there is a relative acceleration between the point particles. The sideways point particles move toward the central one. 
The important point to note is that even though all three of these trajectories are for free falling objects i.e. objects that are “inertial” and have straight line trajectories, as time develops, they are moving toward each other. 
This is clearly a direct violation of Euclid's axioms.
In Euclidean geometry, when we drop the parallelism axiom, we get a curved space. Nevertheless, sufficiently near to any point, we can pretend that the geometry is flat (this is how explained the parallel trajectories in figure \ref{figure}-B). This is true for all so-called Riemannian spaces: they all are locally flat, but the locally straight lines (called geodesics) do not usually remain
parallel. Einstein’s important advance was to see the similarity between Riemannian spaces and gravitational physics. He identified the trajectories of freely falling particles with the geodesics of a curved geometry: they are locally straight since spacetime admits local inertial frames in which those trajectories are straight lines, but globally they do not remain parallel.

%%%%%%%%%%%%%%%%%%%%%%%%%%%%
%%%%%%%%%%%%%%%%%%%%%%%%%%%%

Let us see now how accelerated fields incorporates the physics employed in this example. As we have said, the trajectory of an observer in free fall quantum mechanically is represented by the Hilbert space of an accelerated quantum field with parameter the gravitational acceleration. In this context, the point particle's trajectories illustrated in figure \ref{figure}-B is given by the Hilbert space of $3$-accelerating observers states
\be
\widetilde{\mathtt{a}}_{t_1}^\dagger \,\widetilde{\mathtt{a}}_{t_2}^\dagger\, \widetilde{\mathtt{a}}_{t_3}^\dagger \ket{0_g}.
\ee   
Notice that these states are built using the creation operator of a single field. This entails to the conservation of the number of quanta, a feature we meet at free field theories.
This conclusion cannot be applied to the trajectories shown in figure \ref{figure}-A. The presence of corrections in acceleration imposes the combination of more fields, which entails the creation or annihilation of quanta in the right numbers and with the right length.
In sections \ref{IV} and \ref{V} we will see that the cause of these interactions may be explained by invoking the second law of thermodynamics which pushes the field quantum distribution to approach a black body distribution. 

We should emphasize that spacetime is not the 4-dimensional continuum that appears in general relativity and ordinary quantum field theories;
Space and time arise as excitations of quantum fields. Each quantum carries its own time interval and spatial length, and each field has its own spectrum. If accelerated fields suggests a new language to describe space and time, in this section we argue that the interactions between such fields offer a clear distinction between flat and curved spacetime. Free accelerated field theories are identified with flat spacetime, while interacting accelerated field theories with curved spacetime.  
Given the fact that, according to general relativity, gravity is geometry, then the classification of spacetime in quantum field terms provides a new ground to describe gravity not as a property of Riemannian geometry, but as a quantum fields property.  
Accelerated field theory push physics forward by returning to Einstein’s philosophical roots and extending them in an exciting direction.

We will finish this section by discussing various features of accelerated quantum field theory as applied to free fall observers.

The role of fields is to implement the principle of locality, and
accelerated fields are not a exception. We just need to extend the classical notion of locality, derived from the concept of classical space-time, to momentum space. If we call here the classical locality simply as locality, then, the momentum locality expresses the idea that quantum processes, as described in this section, can be localized in energy and momentum. This is a reasonable consequence of treating energy and momentum as classical entities which appear in the argument of accelerated fields. (We do not claim, against all the experimental facts, that energy and momenta are classical concepts. We just restrict ourselves to a theory that matter fields are not included.)
That non-locality must enter any theory of quantum gravity is guaranteed by the presence of a fundamental length scale. 
On the other hand, one of the properties that quantum field theory incorporates 
into quantum physics is locality.
The framework we propose reconcile that contradiction by linking the generalization of locality to a new understanding of the momentum space. 
An idea which, in different settings, has already been used \cite{PRD84}.

In \cite{C-K2}, free accelerated quantum fields were introduced to show that Unruh effect can be attributed to the nature of space-time.
The relation between inertial and accelerating reference frames is reduced to a correspondence between classical and quantum space-time, expressed by the absence or the presence of accelerated fields, respectively.

This picture cannot survive when gravity is taken into account. In such a system, due to Equivalence principle, all the inertial frames of reference are substituted by free fall ones. As we have shown, in our model, free fall observers entail the quantization of space-time by quantizing their worldline in Minkowski diagram. This means that space-time should be quantized when gravity is not disregarded.
The essential requirement that any theory of gravity must admit local inertial frames (i.e. frames that, at a point, are inertial frames of special relativity), in the model of accelerated fields turns into the prerequisite the interacting theory to acknowledge the existence of a free field theory. Remember that the definition of a free accelerated field is based on the kinematic equation from special relativity.

\section{Source of gravity and Hawking radiation in quantum spacetime}\label{IV}

The results of the previous section suggest that gravity arises as interactions between quanta of accelerated fields, once space and time have emerged from the free theory. 
However we still have not talked about the source of gravity, what is this that cause accelerated fields to interact between each other.
In Newton's description of gravity, the gravitational force is caused by a specific property of material objects: their mass. In Einstein's theory of general relativity, curvature at every point in spacetime is also caused by whatever mass is present (of course it is not the only one, since relativity links mass with energy). 

In our case, too, mass is a key property in determining the effects of gravity. But how can we add mass in a physical system where quantum fields live in momentum space and spacetime is nothing more than the Hilbert space of those fields?

A relativistic invariant way to implement mass is the energy momentum relation $E^2 - p^2 = m^2$.
By construction, the theory of accelerated fields does not convey any information regarding mass. Mass never appears in the equation which defines $\wtG$. 
Since accelerated fields have been established in momentum space, the  consideration of the energy momentum relation modifies the metric of the space, where $\wtG$ is defined, i.e. for any point it holds $(E,p)\rightarrow(\sqrt{p^2+m^2}, p)$.

Below we will investigate the effect of, based on the presence of mass, modified momentum space on the quantum states of $\wtG$. We will not only show the disagreement between an uniform gravitational quantum field built in massless  and in massive momentum space (labeled massless $\wtG$ and massive $\wtG$, respectively) on the definition of the ground state, but also we shall demonstrate that this difference is due to the thermal behavior of massive $\wtG$. This is reminiscent of the Unruh ambiguity inherent to the choice of vacuum for standard quantum fields in Rindler space. In Unruh effect, acceleration is responsible for promoting zero-point quantum field fluctuations to the level of thermal fluctuation. In our case, this role is played by mass.

The approach presented below is based on Ref. \cite{Miloni} where the thermal effect of acceleration was computed by means of the field correlation function.

For the field $\wtG(E,p)$ (from now on $\hbar$ is recovered) satisfying the wave equation (\ref{eq.wem}) we consider the correlation function $
\left\langle \wtG(E_1,p_1)\,\,\wtG(E_2,p_2)\right\rangle_0 $,
computed in the vacuum state.
In this case it holds $\left\langle \widetilde{\mathtt{a}}_{t} \,\,\widetilde{\mathtt{a}}_{t'}\right\rangle_0 = \left\langle \widetilde{\mathtt{a}}_{t}^\dagger \,\,\widetilde{\mathtt{a}}_{t'}^\dagger\right\rangle_0 = \left\langle \widetilde{\mathtt{a}}_{t}^\dagger \,\,\widetilde{\mathtt{a}}_{t'}\right\rangle_0 =0$
and $\left\langle \widetilde{\mathtt{a}}_{t}\,\,\widetilde{\mathtt{a}}_{t'}^\dagger\right\rangle_0 = \delta(t-t')$. 
Therefore we obtain
\be
\left\langle \wtG(E_1,p_1)\,\,\wtG(E_2,p_2)\right\rangle_0 = \frac{\hbar}{\pi}\frac{1}{\Delta E^2 - \Delta p^2}.\label{eq.78-2}
\ee
with $\Delta E=E_2 - E_1$ and $\Delta p=p_2 - p_1$.

In case we measure the vacuum correlation function in massive space, the right part of (\ref{eq.78-2}) can be expressed in another form. 
We can adopt a reparametrization, satisfying automatically the relation $E^2-p^2 = m^2$. As this represents hyperbolic curves, we can use hyperbolic functions and set
\ba
E &=& m \cosh s \label{E1}\\
p &=& m \sinh s \label{p1}
\ea
with $s=\sigma/m$ a variable and $\sigma$ a parameter. 
The induced metric becomes then 
\ba
ds^2 &=& dE^2 - dp^2 \\
     &=& dm^2 - m^2 ds^2
\ea
representing the massive momentum space. There is a horizon at $m=0$ so these coordinates are good for $m > 0$ and $-\infty < \sigma < \infty$. As a consequence, the coordinates $(m,\sigma)$ only cover the patch of momentum space with $E>0$ and $|p|<E$. Thus, a frame of reference embodied with mass is effectively confined to a piece of momentum space and it feels a horizon at $m=0$.

One can calculate the difference $\Delta E^2 - \Delta p^2$ by making use of Eqs. (\ref{E1}) and (\ref{p1})
\be
\Delta E^2 - \Delta p^2 = -4 m^2 \sinh^2\left(\frac{\sigma_2-\sigma_1}{2 m}\right).
\ee 
Then the correlation function $\left\langle \wtPsi(E_1,p_1)\wtPsi(E_2,p_2)\right\rangle_0$ in the vacuum of the uniform gravitational field, defined in massive momentum space, is given by 
\be
\left\langle \wtG(E_1,p_1)\wtG(E_2,p_2)\right\rangle_0 = -\frac{\hbar}{4 \pi m^2} \texttt{csch}^2 \left(\frac{\sigma_2-\sigma_1}{2 m}\right)\label{eq.82} .
\ee

Next, we consider the field correlation function $\left\langle \wtG(0,k)\wtG(0,k+\sigma)\right\rangle$ at a point in (massless) momentum space for a field in equilibrium at temperature $T$. In order to compute this, we impose
\be
\left\langle \widetilde{\mathtt{a}}_{t}^\dagger \,\, \widetilde{\mathtt{a}}_{t'}\right\rangle = \delta(t'-t) n(x_t),\qquad n(x_t)=\left(e^{\frac{x_t}{\widetilde{k} T}}-1\right)^{-1},
\ee
which simply implies that different modes of a thermal field are uncorrelated and that a mode of frequency $x_t$ has an average number of quanta $n(x_t)$.
It is important to mention that due to the dimensions that the Hamiltonian of our system has, the Boltzmann constant has been substituted by $\widetilde{k}_B$, which has dimension length divided by temperature.
Finally we take 
\be
\left\langle \wtG(0,k)\wtG(0,k+\sigma)\right\rangle = -\frac{\hbar}{\pi} \left(\frac{\pi \widetilde{k}_B T }{\widetilde{\hbar}}\right)^2 \texttt{csch}^2 \left(\frac{\pi \widetilde{k}_B T \sigma }{\widetilde{\hbar}}\right),\label{eq.84}
\ee
which, comparing with the correlation function (\ref{eq.82}), we find that they are equivalent for the temperature
\be
T = \frac{\hbar}{2\pi \widetilde{k}_B m}. \label{temper}
\ee 
Assuming the existence of a detector which is specialized in detecting excitations of $\wtG$, the meaning of this result is that this detector in the vacuum,  and defined in massive momentum space, responds as a detector, defined in (massless) momentum space, in a thermal bath at temperature $T=\hbar/2\pi \widetilde{k}_B m$. In other words, a particle with rest mass $m$ radiates quanta of $\wtG$, each one carrying length $x_t=\hbar\widetilde{\omega}$ and time $t=\hbar\widetilde{k}$.

This result raises many issues that should be clarified. 
We will complete this section by discussing and attempting to elucidate some of these. 
First, let us try to give some physical content to the effect that the vacuum state of a massive field $\wtG$ is full of space-time quanta.

The fact that in our construction the vacuum of accelerated fields is unstable to space-time quanta emission in the presence of mass,
should be associated to the conclusion in classical relativistic physics that the Equivalence Principle cannot remove all the effects of gravity in case the system is equipped with mass.

In a uniform gravitational field, gravitation acts on each part of the body equally and this is weightlessness, a condition that also occurs when the gravitational field is zero. Furthermore, the dynamics of the gravitational field, as described in Einstein’s Equations, do not admit solutions that are uniform in space and time. Thus, gravitation becomes apparent through the
non-uniformities in gravitational fields or the tidal forces, as they called. It is these forces, formulated geometrically as space-time curvature, that are regarded as the fundamental manifestation of gravity in general relativity.
In our approach, the tidal effects of gravitation are expressed in quantum terms as quanta in the vacuum state of accelerated quantum fields. The absence of non-uniformities in gravitational field reduces its quantum description to that provided by the accelerated field theory in (massless) momentum space in which the concepts of vacuum and field excitations are well-defined.

Our result, in the first part of this section, does not only show that massless and massive reference frames extract distinct excitation contents from the same field. It also demonstrates that an "observer" in massive reference frame feels a thermal bath of quanta at temperature which is inversely proportional to the mass.   
Unavoidably this makes one think of Hawking effect \cite{Hawking74,Hawking75}. Hawking found that a  Schwarzschild black hole radiates quantum mechanically at a temperature, $T_H = \hbar / 8 \pi G k_B M$, where $M$ is the mass of the black hole.

Probably someone could claim that this comparison would be pointless and unfounded, since involves two unrelated quantum systems.
Hawking discovered that black holes emit particles with a thermal spectrum at a temperature $T_H$ by combining matter quantum fields and classical black hole mechanics.
To be precise, this radiation does not come directly from the black hole itself, but rather is a result of virtual particles being boosted by the black hole's gravitation into becoming real particles.
On the other hand, we attributed a temperature to a massive object just employing accelerated quantum fields and the relativistic energy-momentum relation. 
Classical gravity did not contribute, in any way, to the derivation of the effect.
In a sense, mass promotes vacuum fluctuations of accelerated fields to the level of thermal fluctuations.

Nevertheless,
assuming that Hawking radiation gives a hint on the nature of quantum gravity and since accelerated fields have the ambition to express quantum gravity, $T$ and $T_H$, should be proportional in some limit.
This limit may be the coincidence of the two gravitational objects, namely the black hole mass to satisfy the relation $E^2 - p^2 = m^2$. In doing so,  we implicitly consider black hole as an elementary particle. As it is known this can happen only in Planck scale where quantum gravity dominates. In this scale holds $\widetilde{k}_B/k_B = G$, thus finally we derive
\be
T = 4 \, T_H.
\ee
We find that in Planck scale, black holes, considered just as massive objects, radiate space-time quanta in temperature which is proportional to $T_H$. We should notice that for this result we have not used all the available degrees of freedom of the system of accelerated fields, since we were confined ourselves to $1+1$ dimensions.  
Usually $G$ is treated as a coupling constant. However, in general relativity due to the equivalence principle, $G$ can be understood as a conversion parameter between space-time and energy-momentum space. $G$ in the equation that gives Hawking temperature should be interpreted as conversion parameter as well, which just serves as an auxiliary variable that is needed for dimensional reasons.

But if black hole, as any other massive object, actually radiates space-time quanta, how, in Hawking's analysis, they appeared to radiate energy quanta? This can be explained by the instability in particle production that a gauge/matter field exhibits when it is defined on quantum space-time \cite{C-K2}.

The argument given for Hawking and Unruh effects for the structure of the vacuum near a black hole and acceleration horizon, respectively, applies equally well to accelerated fields vacuum near a mass horizon in momentum space.
As $(E^2 - p^2)\rightarrow \infty$ the temperature is goes down to zero. As the mass horizon, $E=\pm p$, is approached the observer sees a deviating temperature. 

We saw that the momentum vacuum is a thermal state in massive space.
Can we say anything about the entropy associated with this thermal state?
With calculation similar to the one which shows  that, the Minkowski vacuum contains correlations between corresponding modes on either side of the Rindler horizon, one can demonstrate that momentum vacuum accommodates correlations between modes on either side of the mass horizon.
Although the vacuum is a pure quantum state, entanglement implies that its restriction to a localized region is mixed.
 
The corresponding entropy can certainly be infinite (due to the arbitrarily short wavelength fluctuations close to the horizon) in a theory of quantum gravity with no matter fields (remember that accelerated fields is an effort to describe gravity, while it is kept separate
from the matter fields.) 
This result supplements a conclusion in matter quantum field theory, that entropy is infinite provided that no gravity is considered \cite{Jacobson2}.
But how could the horizon entropy ever be finite? In case of matter quantum field theory, it seems that gravity itself should somehow render the entropy finite. We suspect, by analogous arguments, that in accelerated field theory matter fields will be responsible for this.
The investigation of this issue requires the integration of matter fields with accelerated fields, something we intend to do in a subsequent paper.

\section{Mass as a gas of spacetime quanta in thermodynamic equilibrium}\label{V}
In the last  section we have seen that, in the framework of accelerated fields, a massive object manifests its presence by promoting the vacuum fluctuations of the field to the level of thermal radiation. Mathematically, mass has been introduced by modifying the metric of momentum space, the stage where accelerated fields act. This approach allowed us to argue in favor of a correspondence between the, classically defined, curvature of spacetime and the radiation of spacetime quanta, both caused by the presence of mass. However, it did not uncover what is the origin of mass in a context in which space is quantized. On the other hand, the hypothesis of spacetime quanta in connection with statistical mechanics seems sufficient for revealing the nature of mass. In the following I will briefly sketch the method, which largely resembles the derivation of Planck's law from the grand canonical ensemble.

Let the accelerated field $\widetilde{G}$ be confined in a cavity in momentum space which is in thermal equilibrium at temperature $T$.  Adopting the view of previous section, temperature in momentum space is determined in terms of lengths, $T=1/\widetilde{k}_B \, x$. 
Due to the quantization of the field, the cavity with the enclosed accelerated field can be also perceived as gas of spacetime quanta at thermal equilibrium.
Let there be different kind of quanta with the respective numbers $n_t$ and lengths $x_t$ ($t=0$ and $t=\infty$).
Then the total grand canonical partition function of the system is
\ba
\mathcal{Z} &=& \sum_{\{n_i\}}e^{-\frac{1}{\widetilde{k}_B T}\sum_i n_i(x_i-\mu)} \nonumber \\
&=& \prod_i \mathcal{Z}^G_i \label{eq.24}
\ea
where
\be
\mathcal{Z}^G_i = \left(1-e^{\frac{x_i-\mu}{\widetilde{k}_B T}}\right)^{-1}
\ee
the grand canonical partition function for single-quantum state with length $x_i$. $\mu$ is the chemical potential. In the derivation of $\mathcal{Z}^G_i$ in (\ref{eq.24}) we have used the formula from geometric series $1+b+b^2+b^3+\cdots = 1/(1-b)$.
The average number of quanta for that single-quantum state is given by
\be
\left\langle n_i \right\rangle = \left(e^{\frac{x_i-\mu}{\widetilde{k}_B T}} -1\right)^{-1} \label{g-distribution}
\ee
A result that applies for each single-quantum state and thus forms a distribution for the entire state of the system. 

Note that for $\mu=0$, eq.(\ref{g-distribution}) is equivalent to the thermal radiation formula of a massive object, with mass given as a function of temperature.

This result clarifies the picture we should have for massive objects in the framework of accelerated fields. They are collections of spacetime quanta enclosed by a surface in thermal equilibrium which absorbs and re-radiates "length".

The analogy with black body and black body radiation is close and compelling.
Black body is idealized as a cavity full of electromagnetic radiation. According to classical electromagnetism, such a object would absorb all the impinged radiation and could never come to equilibrium with surrounding matter. In thermal terms, it would effectively have a temperature of absolute zero. Something that contradicts observations. Planck and Einstein showed that a black body can reach thermal equilibrium, and thus to radiate, if energy comes in quanta.

Accelerated field theory achieves an equivalent resolution for masses. Classically, both in Newton's law and Einstein's theory, gravity causes every massive object to attract every other massive object. General relativity, in particular, predicts the existence of black holes, regions in which the gravitational effects are so strong that even light can not escape. Nevertheless, the application of quantum theory shows that black holes have nonzero temperature. In this work we have shown that if spacetime is the quantum spacetime that described in accelerated fields theory then, any massive object can reach thermal equilibrium. A massive object, as gravitational black body, absorbs all the, digitized into discrete quanta of length, accelerated field radiation that strikes it. But, to stay in thermal equilibrium, it must emit radiation at the same rate as it absorbs it, so a gravitational black body radiates well. Its emission has a characteristic frequency distribution that depends only on the temperature, or due to (\ref{temper}), on the mass. The analogy with the classical case where the gravitational field that surrounds a massive object depends only its mass is striking.

Just as the Planck's distribution is the unique maximum entropy energy distribution for a gas of photons at thermal equilibrium, so is gravitational black body distribution for a gas of spacetime quanta, but not for energy, for length.
If the gas is not that of gravitational black body's distribution, the second law of thermodynamics guarantees that interactions will cause the spacetime quantum length distribution to change and approach the distribution of a gravitational black body. In such an approach to thermodynamic equilibrium, quanta are created or annihilated in the right numbers and with the right lengths to fill the cavity with the distribution until they reach the equilibrium temperature. Recall that $\mu=0$, which means that the numbers of space-time quanta are not conserved.

The analysis described above can equally be applied to black holes, as massive objects. This should have important implications in the area of black hole thermodynamics. It would be especially interesting to investigate the consequences for the problem of irreversibility.

\section{Summury and discussion}

Quantum gravity remains an outstanding problem of theoretical physics. The bottom line is the physicists are still looking for the nature of the system that should be quantized. In this work we argue that this system should be accelerated fields.

Accelerated field theory provides a theoretical argument that any massive object emits radiation.
The radiation is produced as if emitted by a black body with a temperature inversely proportional to the mass of the object, but contrary to Planck's law, radiation is composed of space-time quanta  and not photons.
In this scheme, gravity appears as a feature of accelerated fields, which is necessary for its consistency. In order to come to this conclusion, we had to give up the classical notion of space-time. The quantization of space-time is imposed by the definition of accelerated fields in momentum space. Thus, gravity continues to be considered a property of space-time.
Depending on the language we use to describe space-time, the definition of gravity is adjusted analogously. 
Classically, where space-time is described as (pseudo-) Riemannian manifold, gravity is represented by its curvature, whereas quantum mechanically, where space-time is expressed in terms of quantum fields, gravity arises as black body radiation. In quantum fields terms the model we propose for quantum gravity is summarized saying that spacetime appears as the Fock space of accelerated fields and gravity create and destroy field quanta, taking us between the different sectors in the Fock space.

The theory presented here on the quantum nature of gravity cannot be the whole story.
Our main result, that massive objects emit black body radiation, was based on the classical form of the energy-momentum relation to serve as source of gravity. However, matter, along with energy, is quantized, so a fully quantized description of gravity requires the incorporation of standard field theory into accelerated quantum fields. The consideration of gravity's source in the same physical form as that of gravity, i.e. as quantum fields, will set us capable of providing a precise formulation of the relationship between space-time fields and matter/gauge fields.

However, we chose the semi-classical approach for two reasons. First, from this point of view, gravity appears to bear a striking resemblance to a well studied effect of standard quantum field theory, Unruh effect.
Second, our findings led us to develop the statistics of space-time quanta, which is actually the Bose-Einstein statistics adapted in the scheme of accelerated quantum fields. It seems to us that the hypothesis of space-time quanta in connection with statistical mechanics is sufficient to define gravity.
From this, we believe that the analysis of many authors \cite{Padmanabhan,Jacobson, Verlinde} on the resemblance between equations of gravity and the laws of thermodynamics can be obtained.  

Although we have considered field theories in $1+1$ dimensions, our results can be extended to physical dimensions. In terms of a global inertial coordinate system
$t,x,y,z$, let us consider the killing field which generates a boost about the origin in the $x$ direction. In this case, the hyperbolic surface $t^2 - x^2 = -1/g^2$ is invariant under translation in $y$ and $z$ direction. Given the fact that the definition of accelerated fields is based on this hyperbolic cylinder surface, in our theory, accelerated
fields become invariant under translations in $y$ and $z$ direction. This clearly reproduces all the results presented above. In the general case, we will have three dimensional surface that will be projected into the planes $(t, x)$, $(t, y)$ and $(t, z)$, thus defining three independent accelerated fields. This reflects the fact that, in our theory, geometry is translated into fields.

\newpage
%%%%%%%%%%%%%%%%%%%%%%%%%%%%%%%%%%%%%%%%%
%\vspace{0.2cm}
\noindent
\textbf{Acknowledgments} --- 
The author would like to thank Lucas C. Céleri for discussions.

\end{document}